\documentclass{emulateapj}

\slugcomment{Accepted for publication in the Astrophysical
 Journal Letters on 21 September 2009.}

\newcommand{\etal}{et~al.}
\newcommand{\eg}{e.g., }
\newcommand{\ie}{i.e., }
\newcommand{\Msun}{M_{\odot}}
\newcommand{\Rsun}{R_{\odot}}
\newcommand{\Menv}{M_{\rm env}}
\newcommand{\Ms}{M_{\rm preSN}}
\newcommand{\Rs}{R_{\rm preSN}}

\newcommand{\Cofs}{$^{56}$Co}
\newcommand{\Nifs}{$^{56}$Ni}
\newcommand{\Mni}{M{\rm (^{56}Ni)}}
\newcommand{\Mms}{M_{\rm ZAMS}}
\newcommand{\Mej}{M_{\rm ej}}

\newcommand{\SN}{SNLS-04D2dc}
\newcommand{\Ebvh}{E_{B-V, {\rm host}}}
\newcommand{\Ebvg}{E_{B-V, {\rm Gal}}}
\def\gsim{\mathrel{\rlap{\lower 4pt \hbox{\hskip 1pt $\sim$}}\raise 1pt
\hbox {$>$}}}
\def\lsim{\mathrel{\rlap{\lower 4pt \hbox{\hskip 1pt $\sim$}}\raise 1pt
\hbox {$<$}}}

\begin{document}

\title{Properties of Type II Plateau Supernova SNLS-04D2dc: Multicolor
Light Curves of Shock Breakout and Plateau}

\author{
 N.~Tominaga\altaffilmark{1,2,3},
 S.~Blinnikov\altaffilmark{4,2},
 P.~Baklanov\altaffilmark{4,5},
 T.~Morokuma\altaffilmark{3,6},
 K.~Nomoto\altaffilmark{2,7},
 T.~Suzuki\altaffilmark{7}
 }

\altaffiltext{1}{Department of Physics, Faculty of Science and
Engineering, Konan University, 8-9-1 Okamoto,
Kobe, Hyogo 658-8501, Japan; tominaga@konan-u.ac.jp}
\altaffiltext{2}{Institute for the Physics and Mathematics of the
Universe, University of Tokyo, 5-1-5 Kashiwanoha, Kashiwa, Chiba
277-8569, Japan}
\altaffiltext{3}{Optical and Infrared Astronomy Division, National
Astronomical Observatory, 2-21-1 Osawa, Mitaka, Tokyo 181-8588, Japan; 
tomoki.morokuma@nao.ac.jp}
\altaffiltext{4}{Institute for Theoretical and  Experimental Physics (ITEP),
Moscow 117218, Russia; sergei.blinnikov@itep.ru, baklanovp@gmail.com}
\altaffiltext{5}{Max-Planck-Institute for Astrophysics,
Karl-Schwarzschild-Str. 1, 85741 Garching, Germany}
\altaffiltext{6}{Research Fellow of the Japan Society for the Promotion of Science}
\altaffiltext{7}{Department of Astronomy, School of Science,
University of Tokyo, Bunkyo-ku, Tokyo 113-0033, Japan;
nomoto@astron.s.u-tokyo.ac.jp, suzuki@astron.s.u-tokyo.ac.jp}

\begin{abstract}

 Shock breakout is the brightest radiative phenomenon in a Type II
 supernova (SN). 
 Although it was predicted to be bright, the direct observation
 is difficult due to the short duration and X-ray/ultraviolet-peaked
 spectra. First entire observations of the shock breakouts of Type
 II Plateau SNe (SNe IIP) were reported
 in 2008 by ultraviolet and optical observations by the {\it GALEX}
 satellite and supernova legacy survey (SNLS), named \SN\ and
 SNLS-06D1jd. We
 present multicolor light curves of a SN IIP, including the shock breakout
 and plateau, calculated with a multigroup radiation hydrodynamical code
 {\sc STELLA} and an evolutionary progenitor model.
 The synthetic multicolor light curves reproduce
 well the observations of \SN. This is the first study to
 reproduce the ultraviolet light curve of the shock breakout and the optical
 light curve of the plateau consistently. We conclude that
 \SN\ is the explosion with a canonical explosion energy
 $1.2\times10^{51}$ ergs and that its progenitor is a star with a
 zero-age main-sequence mass $20\Msun$ and a presupernova radius
 $800\Rsun$. The model demonstrates that the peak apparent $B$-band
 magnitude of the shock breakout would be $m_{\rm B}\sim26.4$~mag if a
 SN being identical to \SN\ occurs at a redshift $z=1$, which can be reached by
 8m-class telescopes. The result evidences that the shock breakout has a
 great potential to detect SNe IIP at $z\gsim1$.
\end{abstract}

\keywords{shock waves --- radiative transfer --- supernovae: general ---
supernovae: individual (SNLS-04D2dc) --- stars: evolution}

\section{INTRODUCTION}
\label{sec:intro}

In a supernova (SN) explosion, an outward shockwave formed in an inner
layer propagates through the stellar envelope. When the shock
emerges from the stellar surface, a hot fire ball suddenly appears and
flashes in soft X-ray or ultraviolet (UV). The flash has 
a quasi-blackbody spectrum ($T\sim10^6-10^5$~K) and lasts a few 
seconds to $\sim1$~days, depending on an ejecta mass $\Mej$,
explosion energy $E$, and presupernova radius $\Rs$ (\eg
\citealt{mat99}). The phenomenon is called ``shock breakout'' having
been theoretically predicted (\eg \citealt{kle78}).

Owing to the short duration and X-ray/UV-peaked spectra, it is
difficult to observe directly the shock breakout. The first detection of shock
breakout was reported for nearby Type II-peculiar SN~1987A
\citep{cat87,ham88}, but it was only a detection of a rapid decline
presumed to be a shock breakout tail. Although the detections of shock
breakout tails were occasionally reported for Type IIb SN~1993J \citep{ric94}
and Type Ib SN~1999ex \citep{str02}, there were no observations of shock
breakouts from the rising part or with the X-ray or UV light.

First entire observations of the shock breakouts were reported in
2008. A first example is an X-ray observation of Type Ib SN~2008D
which fortunately appeared in the same
galaxy as SN~2007uy during the observation by the {\it Swift} satellite
(\citealt{sod08,maz08,mod08,mal09}). Second examples are Type II plateau SNe (SNe IIP) \SN\
($z=0.185$, \citealt{sch08,gez08}) and SNLS-06D1jd ($z=0.324$,
\citealt{gez08}). They were caught coincidentally in the UV Deep Imaging
Survey by the {\it GALEX} satellite \citep{mor05,mor07} at the location
where supernova legacy survey (SNLS, \citealt{ast06}) found SN
candidates. 

Since SNe IIP are the most common among core-collapse
types of SNe (\eg \citealt{man08}) and their shock breakouts are suggested
to be so bright as to be detected even at $z\gsim1$
(\eg \citealt{chu00}), it is important to develop a way
to derive SN properties from the shock breakouts.
Therefore, we calculate multicolor
light curves (LCs), including a shock breakout and plateau, of SN IIP
based on an evolutionary progenitor model with a multigroup radiation
hydrodynamics code {\sc STELLA} \citep{bli98,bli00,bli06}. The synthetic
LCs are compared with the multicolor observations of SN IIP
\SN\ (${\rm R.A.=10^{h}00^m16.7^s}$, ${\rm decl.=+02^\circ12'18.52''}$
[J2000.0]).\footnote{We focus on \SN\ because SNLS-06D1jd has sparse UV
observations with relatively low signal-to-noise ratio.} We
first present a multicolor LC model reproducing well the shock 
breakout and plateau consistently and constrain SN and progenitor
properties. Furthermore, based on the multicolor LC model, we present
an apparent $B$-band light curve of a shock breakout of a SN
being identical to \SN\ at $z=1$.

In \S~\ref{sec:model}, the applied models and the radiation
hydrodynamics calculations are described. In \S~\ref{sec:result},
the multicolor LCs of \SN\ are compared with the synthetic
LCs. In \S~\ref{sec:discuss}, conclusion and discussion are
presented.

\section{Methods \& Models}
\label{sec:model}

We apply the multigroup spherical radiation hydrodynamics code {\sc STELLA}
\citep{bli98,bli00,bli06}. {\sc STELLA} adopts variable Eddington
factors, a gray transfer of $\gamma$-ray from radioactive nuclei, LTE
ionization states, and a multigroup expansion opacity and solves the
time-dependent equations implicitly for the angular moments of intensity
averaged over fixed frequency bands (for details, see
\citealt{bli06} and references therein).
Multigroup radiative transfer is coupled with hydrodynamics, which
enables to acquire the spectral energy distributions (SEDs) self-consistently. The
color temperature of a SN is estimated from a black-body fitting of
the SED. In this Letter we adopt 100
frequency bins dividing logarithmically from $\lambda=1$~\AA\ 
to $5\times10^4$~\AA; the large number of frequency bins
allows to describe accurately a non-equilibrium continuum radiation.

A progenitor model is a non-rotating solar-metallicity star constructed
by a stellar evolution calculation \citep{ume05}. The calculation includes a 
metallicity-dependent mass loss \citep{kud00} and thus the presupernova
model has a self-consistent $\Rs$, luminosity, temperature, envelope
mass, and total mass.\footnote{We note that the presupernova progenitor structure
depends on the treatments of physics, \eg rotation, mass loss, mixing length, and
overshooting (\eg \citealt{lim06}).} Since the shock breakout and plateau
depend on $\Mej$, $E$, and $\Rs$ (\eg \citealt{eas94,mat99}), our
calculation achieves self-consistent multicolor LCs from the shock
breakout to plateau and tail. In this Letter we present a SN explosion
of a star with a zero-age main-sequence mass $\Mms=20\Msun$ having a presupernova mass
$\Ms=18.4\Msun$, H envelope mass $\Menv=13.4\Msun$, and presupernova
radius $\Rs=800\Rsun$. An extensive investigation will be presented in
a forthcoming paper.

\section{Comparisons with observations}
\label{sec:result}

\cite{sch08} and \cite{gez08} found a UV brightening at the \SN\
position which lasts several days from $\sim15$ days before the first
SNLS observation. Although the optical observation of the shock
breakout is not available, the UV-optical LCs of the shock breakout and
plateau can be compared with the synthetic multicolor LCs. We assume the
date of shock breakout to be 2453062.2 JD ($t=0$) and compare the model
and observations with reference to the date. In this Letter the epochs
are described in the observer frame.

\begin{figure}
\plotone{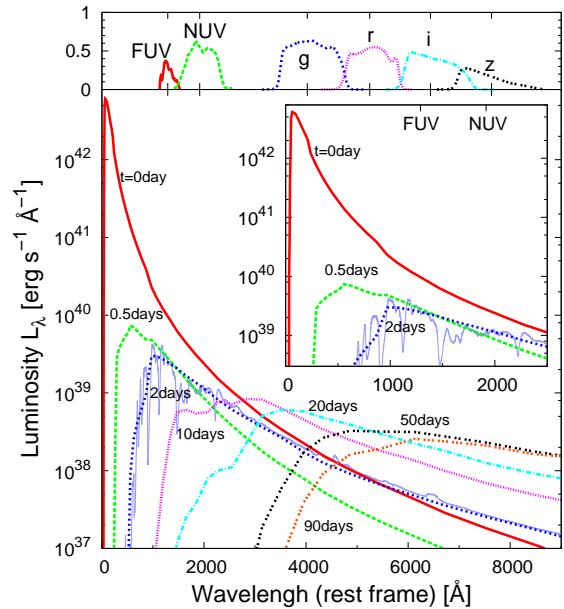} 
\caption{{\it Top}: Sensitivity curves of multicolor bands ({\it
 red}: FUV-band, {\it green}: NUV-band, {\it blue}: $g$-band, {\it
 magenta}: $r$-band, {\it cyan}: $i$-band, and {\it black}: $z$-band). For
 illustration purpose, each filter band is blueshifted to the rest frame
 to compensate for $z=0.185$.
{\it Bottom}: Evolution of intrinsic SEDs of a SN IIP model with
 $\Mms=20\Msun$ and $E=1.2\times10^{51}$~ergs at $t=0$~day
 ({\it red}), $0.5$~days ({\it green}), $2$~days ({\it blue}), $10$~days
 ({\it magenta}), $20$~days ({\it cyan}), $50$~days ({\it
 black}), and $90$~days ({\it orange}). A synthetic non-LTE spectrum is
 also shown ({\it violet}, \citealt{gez08}). The inset
 enlarges the UV emission at $t=0,~0.5,$ and $2$~days and the non-LTE
 spectrum.
}
\label{fig:SED}
\end{figure}

The multigroup spherical radiation hydrodynamics calculation provides
wavelength- and time-dependent fluxes at the SN surface. When a SN is observed
from a given direction, lights from different parts of the SN surface
are radiated at different time and at different radii (for details,
see \citealt{kle78,ims81,ens92,bli02,bli03}). Thus, we take into account
a light travel time correction and limb darkening in the Eddington
approximation \citep{kle78}. Figure~\ref{fig:SED} shows corrected
wavelength-dependent luminosities $L_\lambda$ for a SN IIP model with
$\Mms=20\Msun$ and $E=1.2\times10^{51}$~ergs. 

The SED at $t=2$~days is compared with
synthetic non-LTE spectrum ($55.6$~hr, \citealt{gez08}; see also
\citealt{des05}) which gives similar optical color. Although the epochs
are different because of adopting different progenitor models, the UV SED
and spectrum derived from the independent calculations are distinctly
consistent. The consistency justifies both theoretical calculations.

In order to predict multicolor observations from the multicolor
theoretical model,
the model lights are diluted,\footnote{The distance is derived with
the five-year result of {\it Wilkinson Microwave Anisotropy Probe}
\citep{kom09}. } redshifted, reddened, and then convolved with the
sensitivities of the satellite and telescope ({\it GALEX}:
\citealt{mor05,mor07}, the MegaPrime/MegaCam on the
Canada-France-Hawaii Telescope (CFHT) for SNLS: \citealt{ast06}).
For illustration purpose, the sensitivity curves blueshifted to the rest
frame to compensate for $z=0.185$ are shown in the top panel of
Figure~\ref{fig:SED}. In this Letter the bands are described in the
observer frame.

\subsection{Ultraviolet light curves of shock breakout}
\label{sec:UV}

\begin{figure}
\plotone{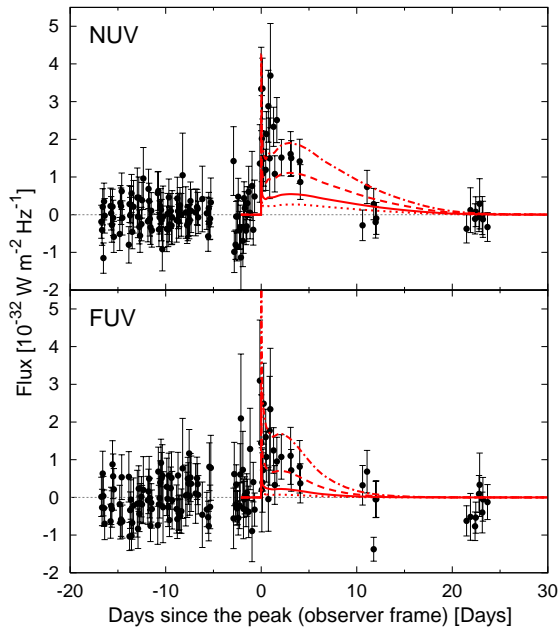} 
\caption{Comparison between the NUV ({\it top}) and FUV ({\it bottom}) 
 observations ({\it points}, \SN, \citealt{sch08})
 and the SN IIP model without the host galaxy extinction ({\it dot-dashed line}) and
 reddened for the host galaxy extinction with $\Ebvh=0.06$~mag ({\it dashed
 line}), $0.14$~mag ({\it solid
 line}), and $0.22$~mag ({\it dotted line}). 
}
\label{fig:SNLSuv}
\end{figure}

The UV light allows for direct observations of the shock breakout but
is strongly reduced by extinction. Thus it is crucial to estimate
correctly the extinction in the host galaxy and our Galaxy. While the color excess of our
Galaxy $\Ebvg$ is taken from Schlegel \etal\ (1998, $\Ebvg=0.02$~mag), it is
difficult to estimate the color excess and thus extinction of the
host galaxy. \cite{sch08} estimates the color excess of the host galaxy
$\Ebvh$ at the SN location from the Balmer decrement as
$\Ebvh=0.14$~mag. However, they caution that the uncertainty of the
total extinction is as much as a factor of two and
further note that the estimate from the empirical relation of SNe IIP
\citep{nug06} is consistent with both of $\Ebvh=0.14$~mag and $0$.

Assuming the Small Magellanic Cloud (SMC) reddening law for the host
galaxy \citep{pei92} and $\Ebvh=0.14$~mag, the total extinction at 
the effective wavelengths of the far
and near UV (FUV and NUV) filters of the {\it GALEX} satellite are as
large as $A_{\rm FUV}=2.38$~mag and $A_{\rm NUV}=1.51$~mag, respectively. 
Hereafter, we call these values ``standard'' extinction.
Although the values are slightly different from \cite{sch08}, 
the total extinction integrated over each band depends on the
intrinsic spectrum and varies with time as the spectrum changes.
The variations of extinction in the UV bands are relatively
large; for example, $\sim0.3$~mag in the FUV band and $\sim0.1$~mag in the
NUV band from $t=0$ to 20 days for the SN IIP model with $\Mms=20\Msun$
and $E=1.2\times10^{51}$~ergs.

Because of the large uncertainty, we assume several values for the color
excess of the host galaxy as follows: $\Ebvh=0$~mag referring to the
case of no extinction in the host galaxy, $\Ebvh=0.06$~mag giving half
of the standard extinction in the NUV band, $\Ebvh=0.14$~mag being the
standard extinction, and $\Ebvh=0.22$~mag giving double of the
standard extinction in the NUV band, which lead to
$(A_{\rm NUV}, A_{\rm FUV})=(0.18~{\rm mag},~0.15~{\rm mag})$,  
$(0.75~{\rm mag},~1.10~{\rm mag})$, $(1.51~{\rm mag},~2.38~{\rm mag})$, 
and $(2.27~{\rm mag},~3.65~{\rm mag})$, respectively. Here, we assume
the SMC reddening law for the host galaxy.

Figure~\ref{fig:SNLSuv} shows comparisons of UV LCs with the model with 
$\Mms=20\Msun$ and $E=1.2\times10^{51}$~ergs. The model LCs are
consistent with the observations within the uncertainty, while they are
slightly fainter than the observations for $\Ebvh=0.14$~mag. 
The explosion energy of SNLS-04D2dc is consistent
with the canonical value of the explosion energies of core-collapse SNe
[\eg SN~1987A: $E=(1.1\pm0.3)\times10^{51}$~ergs, \citealt{bli00}]. 
Although the \Nifs-\Cofs\ radioactive decay does not contribute to the
shock breakout, we expediently assume a canonical \Nifs\ ejection
without mixing to the envelope [the ejected \Nifs\ mass
$\Mni=0.07\Msun$, \eg SN~1987A: \citealt{bli00}], and thus
$\Mej=16.9\Msun$ to yield $0.07\Msun$ of \Nifs.

\begin{figure}
\plotone{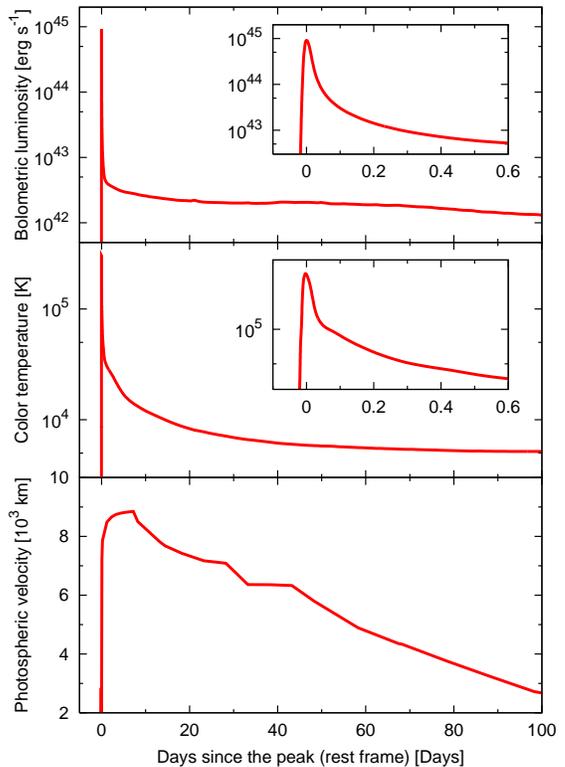} 
\caption{Bolometric LC ({\it top}), color temperature evolution ({\it
 middle}), and photospheric velocity evolution ({\it bottom}) of the SN
 IIP model ({\it lines}). The insets in the top and middle panels
 enlarge the phase of shock breakout. 
}
\label{fig:model}
\end{figure}

The second peak in the NUV LC at $t\sim3$~days is reproduced by the
model and explained by the shift of the peak wavelength as
\cite{sch08} and \cite{gez08} suggested. The bolometric LC and the evolution of color
temperature are shown in Figure~\ref{fig:model}. Figure~\ref{fig:model}
also shows the velocity evolution of photosphere defined as a position
where the radiation and gas are decoupled. Although the bolometric
luminosity declines monotonically after the shock breakout, the
radiation energy in the NUV band increases with time
because the peak wavelength shifts long (Fig.~\ref{fig:SED}).
After the NUV second peak, the color temperature
decreases further and the peak wavelength shifts to the optical bands. 
The shift is caused by not only the decreasing temperature of the SN
ejecta but also an enhancement of the metal absorption lines due to the
low temperature.
As a result, the UV luminosity declines monotonically after the NUV second
peak. The model also predicts a second peak in the FUV band but the
brightening is obscured because of the low signal-to-noise ratio.

\subsection{Optical light curves at plateau stage}

\begin{figure}
\plotone{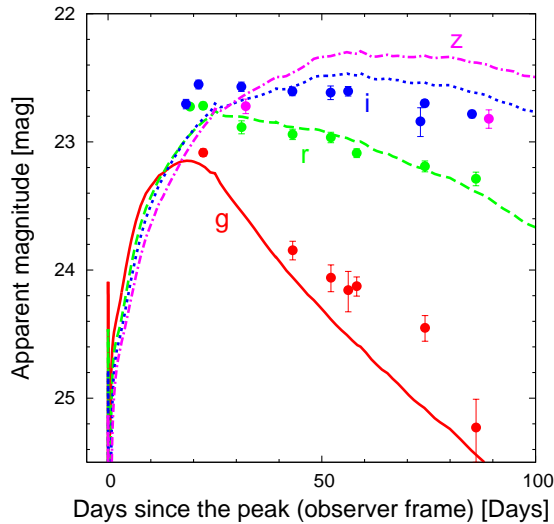} 
\caption{Comparison between the SNLS observations
 ({\it points}, SNLS-04D2dc, \citealt{sch08})
 and the SN IIP model reddened for the host galaxy extinction
 with $\Ebvh=0.14$~mag ({\it lines}) ({\it red}: $g$-band, {\it green}:
 $r$-band, {\it blue}:  $i$-band, {\it magenta}: $z$-band in AB
 magnitude system).
}
\label{fig:SNLSopt}
\end{figure}

Thanks to the multigroup radiation hydrodynamics calculations,
subsequent evolutions of multicolor lights are obtained and compared with
the SNLS optical observations.
Figure~\ref{fig:SNLSopt} shows comparisons of the $g$-, $r$-, $i$-, and
$z$-band LCs. Here, we adopt $\Ebvh=0.14$~mag and the SMC reddening
law for the host galaxy. The total extinction at the effective
wavelengths of the $g$-, $r$-, $i$-, and $z$-band filters of the
MegaPrime/MegaCam on CFHT are as large as $A_g=0.64$~mag,
$A_r=0.47$~mag, $A_i=0.36$~mag, and $A_z=0.28$~mag, respectively. 

As SED peaks in the $g$-band at $t\sim20$~days
(Fig.~\ref{fig:SED}), the $g$-band LC peaks at $t\sim20$~days. After
this epoch, the SED becomes red with time and the blue edge of the
SED enters in the $g$-band. Thus, the $g$-band LC declines more rapidly than
the other optical-band LCs. On the other hand, the $r$-band LC declines
gradually and the decline rate slightly changes at $t\sim55$~days
when the SED peak enters in the $r$-band. The $i$- and $z$-band LCs
brighten by $t\sim60$~days due to the shift of the peak wavelength.

Although the overall multicolor LCs are well reproduced by
the model, the $z$-band model LC is brighter than the observation at
$t\sim90$~days. The discrepancy would be improved if we add more physics to {\sc
STELLA}. {\sc STELLA} currently has around $1.5\times10^5$ spectral
lines from \cite{kur95} and \cite{ver96} and a rather poor line list in
near infrared, while tens of millions of spectral lines from large
Kurucz lists are being included (E. Sorokina, private
communication). The expansion of line lists, together with taking into
account non-equilibrium effects in an important coolant like
\ion{Ca}{2}, may influence the $z$-band flux appreciably \citep{kas06}.
 
No mixing of \Nifs\ leads to no contribution from the radioactive heating to
the LC at $t\lsim100$~days, while the radioactive decay could power the
LC from the early phases if \Nifs\ is mixed to the envelope. Because of
the lack of observations at $t>100$~days, $\Mni$ of \SN\ cannot be
constrained. We note that the available observations can be reproduced
equally well even if no \Nifs\ is ejected.

\begin{figure}
\plotone{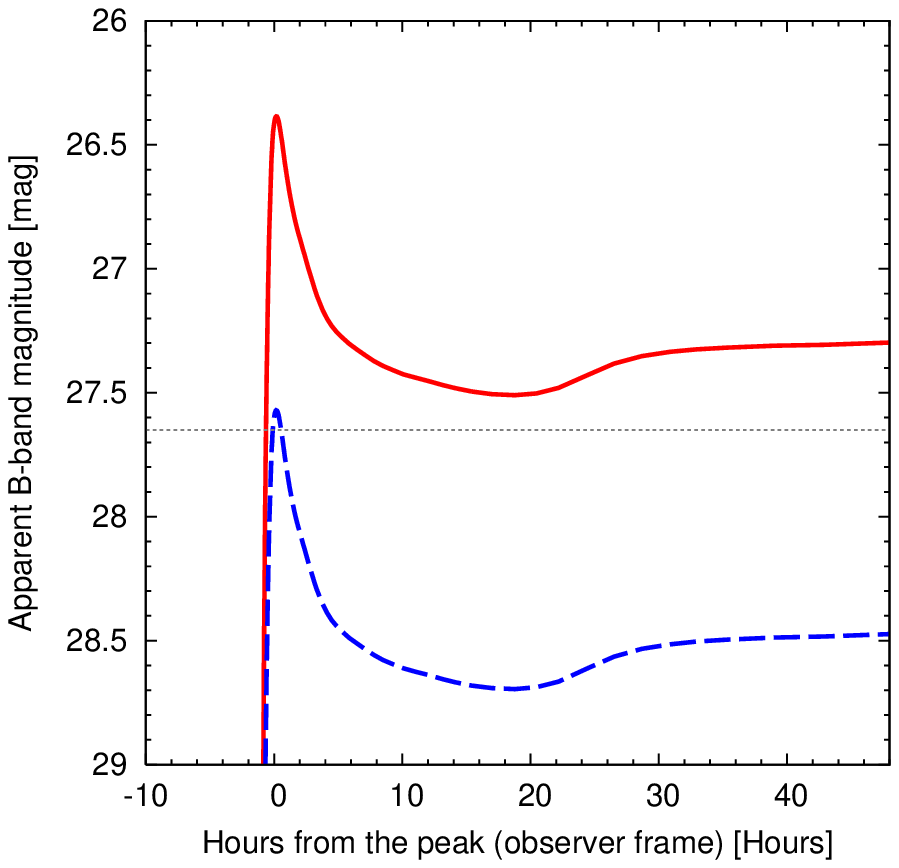}
\caption{Apparent $B$-band light curve of a shock breakout in AB magnitude system
 for a SN being identical to SNLS-04D2dc at $z=1$. We assume
 $\Ebvg=0.02$~mag, $\Ebvh=0$ ({\it solid line}) and $0.14$~mag ({\it
 dashed line}), and the SMC reddening law for the host galaxy. 
 A limiting magnitude for a $4\sigma$ detection in a hour of
 integration with SUBARU/Suprime-Cam \citep{miy02} is
 also shown ({\it dotted line}), which is calculated with Subaru
 Imaging Exposure Time Calculator
 (http://www.naoj.org/cgi-bin/spcam\_tmp.cgi) assuming 
 $0.5''$ seeing, $1.5''$ aperture, and 3 days from New Moon.
}
\label{fig:z1}
\end{figure}

\section{CONCLUSIONS \& DISCUSSION}
\label{sec:discuss} 

We present a multicolor LC model with the multigroup radiation hydrodynamical code
{\sc STELLA}. The model reproduces the multicolor
UV-optical LCs of the shock breakout and plateau of \SN\
consistently. Since the shock breakout is more sensitive to $E$ and
$\Rs$ than the plateau \citep{eas94,mat99},
the SN and progenitor properties are constrained more tightly than
only with the plateau observation.
We clarify the properties of \SN: the progenitor is a
star with $\Mms=20\Msun$, $\Ms=18.4\Msun$, and $\Rs=800\Rsun$, and the
SN has the canonical explosion energy $E=1.2\times10^{51}$~ergs. 
Because of no observations at $t>100$~days, we cannot constrain
$\Mni$ of \SN, while we expediently adopt $\Mej=16.9\Msun$ and
$\Mni=0.07\Msun$.

The second peak of the NUV LC is reproduced by the shift of the peak
wavelength, in spite of the monotonic decline of the bolometric LC.
Although the signal-to-noise ratio of the FUV LC is quite low, the model is also in
an agreement with the FUV LC. The consistency supports the origin of the
NUV second peak. When only the monotonic light curve is obtained, only
two observational quantities, \ie flux and duration, are available
and the three properties characterizing the shock breakout, $\Mej$,
$\Rs$, and $E$, are degenerated. The multicolor observation 
enables us to constrain the color temperature and thus to resolve the
degeneracy. In order to clarify the SN properties in detail, it is
required to employ the multigroup radiation hydrodynamical calculations
because the spectra of the shock breakout slightly deviate from the
blackbody.

The successful reproduction of UV-optical LCs of SNLS-04D2dc and the
consistency with the non-LTE spectral calculation
justify the multigroup radiation
hydrodynamics calculation with {\sc STELLA} and indicate that the
calculations are capable of predicting the multicolor observations of
the shock breakout and plateau of SNe IIP. If a SN being
identical to \SN\ takes place at $z=1$ in the direction with
$\Ebvg=0.02$~mag, the SN brightness will be 
$m_{\rm B}\sim26.4$~mag without the host galaxy extinction or
$m_{\rm B}\sim27.6$~mag with the
host galaxy extinction with $\Ebvh=0.14$~mag and the SMC reddening law
(Fig.~\ref{fig:z1}). It is bright enough to be detected with 8m-class
telescopes. The result envisages that
future deep and wide surveys by, \eg SkyMapper, Panoramic Survey
Telescope and Rapid Response System (Pan-STARRS), Large Synoptic Survey
Telescope (LSST), and SUBARU/Hyper Suprime-Camera (HSC), will find a
large number of shock breakouts. The large sample will import
totally-new knowledge about cosmic evolution histories, \eg a
core-collapse SN rate and a star formation rate, at $z\gsim1$. 
Therefore, although the detections of Type IIn SNe at $z>2$ are
recently reported \citep{coo09}, we still emphasize a great potential
of shock breakouts for direct detection of the most common
core-collapse SNe, SNe IIP, at $z\gsim1$. 

Finally, we point out that the estimate of the host galaxy extinction 
is crucial to estimate the intrinsic luminosity of shock breakout.
This is because the shock breakout of SN IIP emits the radiation energy
mainly in UV easily reduced by the interstellar extinction. With the
use of the sensitiveness, if the observations of shock breakout,
plateau, and tail are available with high signal-to-noise ratios,
finding a set of a SN model and extinction, which consistently
reproduces the observations, might provide a new constraint on the
host galaxy extinction.

\acknowledgments

We thank Kevin Schawinski and the co-authors of \cite{sch08} paper and
Luc Dessart and the co-authors of \cite{gez08} paper for kindly
providing the data of FUV observations and the data of synthetic
non-LTE spectra of the shock breakout, respectively. S.B. thanks
Katsuhiko Sato and Hitoshi Murayama for hospitality at University of
Tokyo during his visits in 2006-2009.
N.T. and T.M. have been supported by the JSPS (Japan Society for the
Promotion of Science) Research Fellowship for Young Scientists.
The work of S.B. and P.B. in Russia is supported partly by the grant RFBR
07-02-00830-a,  by Scientific School Foundation under grants
2977.2008.2, 3884.2008.2, and in Germany by MPA guest program.
This research has been supported in part by World Premier
International Research Center Initiative, MEXT,
Japan, and by the Grant-in-Aid for Scientific Research of the JSPS
(18104003, 20540226, 21840055) and MEXT (19047004, 20040004).

\end{document}